\documentclass[conference, 10pt]{IEEEtran}

\usepackage{epsfig}
\usepackage{flushend}
\usepackage{fixltx2e}
\usepackage{url}
\usepackage{authblk}
\usepackage{amsmath}
\usepackage[utf8]{inputenc}
\usepackage{amssymb}
\usepackage{graphicx}
\usepackage{subfig}
\begin{document}

\title{OnionBots: Subverting Privacy Infrastructure\\ for Cyber Attacks}

\author{Amirali Sanatinia\\ Northeastern University \\amirali@ccs.neu.edu
        \and Guevara Noubir\\ Northeastern University \\noubir@ccs.neu.edu}

\maketitle

\begin{abstract}
Over the last decade botnets survived by adopting a sequence of increasingly sophisticated strategies to evade detection and take overs, and to monetize their infrastructure. At the same time, the success of privacy infrastructures such as Tor opened the door to illegal activities, including botnets, ransomware, and a marketplace for drugs and contraband. We contend that the next waves of botnets will extensively subvert privacy infrastructure and cryptographic mechanisms. 
In this work we propose to preemptively investigate the design and mitigation of such botnets. We first, introduce OnionBots, what we believe will be the next generation of resilient, stealthy botnets. OnionBots use privacy infrastructures for cyber attacks by completely decoupling their operation from the infected host IP address and by carrying traffic that does not leak information about its source, destination, and nature. Such bots live symbiotically within the privacy infrastructures to evade detection, measurement, scale estimation, observation, and in general all IP-based current mitigation techniques. Furthermore, we show that with an adequate self-healing network maintenance scheme, that is simple to implement, OnionBots achieve a low diameter and a low degree and are robust to partitioning under node deletions. We developed a mitigation technique, called SOAP, that neutralizes the nodes of the basic OnionBots. We also outline and discuss a set of techniques that can enable subsequent waves of Super OnionBots. In light of the potential of such botnets, we believe that the research community should proactively develop detection and mitigation methods to thwart OnionBots, potentially making adjustments to privacy infrastructure.
\end{abstract}

\section{Introduction}\label{sec:intro}

Over the last decade botnets rose to be a serious security threat. They are routinely used for denial of service attacks, spam, click frauds, and other malicious activities~\cite{bot:survey}. Both the research and industry communities invested a significant effort analysing, developing countermeasures, and products to effectively detect, cripple, and neutralize botnets. While some countermeasures operate on user computers,  most are deployed at the ISP and enterprise levels. Many botnets were successfully neutralized by shutting down or hijacking their Command and Control (C\&C) servers, communications channels (e.g., IRC), reverse engineering the algorithm used for the domain name generation (DGA) and preemptively blocking the access to these domains~\cite{Silva2013378}. Such mitigation techniques exploit the fact that most current botnets rely on primitive communication architectures and C\&C mechanisms. This forced botnet developers to continuously adapt raising the level of sophistication of their design from the early static and centralized IRC or fixed servers' IP addresses to more sophisticated fast-fluxing~\cite{fastflux} and even straightforward use of Tor hidden services~\cite{Zeus64},~\cite{ChewBacca}.

In this paper, we are interested in investigating the next level of this arm-race. We contend that the next wave of botnets' sophistication will rely on subverting privacy infrastructure and a non-trivial use of cryptographic mechanisms. The Tor project was very successful in building an infrastructure that protects users identity over the Internet and allowing one to host Internet servers without revealing her or his location using the Tor hidden services feature. Evidence of our predictions can be found in the malicious use of hidden services for hosting the infamous silk road~\cite{Christin:2013:silkroad}, instances of the Zeus~\cite{Zeus64} botnet, and the hosting of the CryptoLocker ransomware C\&C server~\cite{cryptolocker}. Interestingly, CryptoLocker combines Tor with the use of another privacy ``infrastructure'', bitcoin the crypto currency, for the ransom payment. The combination of Tor and bitcoin make it possible today to blackmail Internet users, anonymously be paid, and get away with it. 

The current use of Tor and crypto-mechanisms in botnets is still in its infancy stages. Only hosting the C\&C server as a hidden service still allows the detection, identification, and crippling of the communication between the bots. Recent research demonstrated that it is possible to successfully deny access to a single or few \texttt{.onion} server~\cite{trawltor}. To assess the threat of crypto-based botnets, we advocate a preemptive analysis, understanding of their potential and limitations, and the development of mitigation techniques. 

In this paper, we present the design of a first generation of non-trivial OnionBots. In this Basic OnionBot, communication is exclusively carried out through hidden services. No bot (not even the C\&C) knows the IP address of any of the other bots. At any instant, a given bot is only aware of the \textit{temporary} \texttt{.onion} address of a very small (constant) number of bots. Bots relay packets but cannot distinguish the traffic based on their source, destination, or nature. At the same time, the bot master is able to access and control any bot, anytime, without revealing his identity. We show that this design is resilient to current mitigations and analysis techniques from botnet mapping, hijacking, to even assessing the size of the botnet. We also show that the proposed Neighbors-of-Neighbor graph maintenance algorithm achieves a low diameter, degree, and high resiliency and repair in the event of a take-down (e.g., Tor DoSing or node capture/cleanup) of a significant fraction of the botnet nodes. Since our goal is to preemptively prevent the emergence of OnionBots, we also propose a novel mitigation technique against the Basic OnionBots. This technique exploits the same stealthy features of the OnionBot (namely peers not knowing each other's identities) to neutralize the bots. The technique called SOAP, gradually surrounds the bots by clones (or sybils) until the whole botnet is fully contained.

We believe that the Basic OnionBot is only the prelude to waves of increasingly sophisticated bots subverting privacy infrastructure such as Tor. We outline and discuss an additional set of techniques that fully exploits the decoupling between IP addresses and bots, enabling multi-bot per host, proofs of work mechanisms and other techniques that can potentially overcome the SOAP mitigation. Our goal is to draw the attention of the community to the potential of OnionBots and develop preemptive measures to contain them and ideally prevent their occurence.

Our contributions are summarized as follows:

\begin{itemize}

\item A novel reference design for a OnionBotnet whose command, communication, and management are fully anonymized within the Tor privacy infrastructure.

\item A communication topology with repair mechanisms that minimizes the nodes' degree, graph diameter, and maximizes resiliency.

\item A performance evaluation and discussion of resiliency to various takedown attacks such as simultaneous denial of service attacks against multiple \texttt{.onion} botnet nodes. 

\item A Sybil mitigation technique, SOAP, that neutralizes each bot by surrounding it by clones. 

\item An outline and discussion of a set of techniques that can enable subsequent waves of Super OnionBots.

\end{itemize}

We first survey the current state of botnet design and mitigation techniques in Section~\ref{s:current}, followed by a review of the key features of the Tor privacy infrastructure in Section~\ref{s:tor}. In Section~\ref{s:onionbot}, we present our proposed reference design for an OnionBotnet. We evaluate the resiliency and performance of the OnionBotnet, using several metrics in Section~\ref{s:eval}. We investigate potential mechanisms to prevent the rise of such botnets in Section~\ref{s:mitigation}. We finally discuss how the attackers would evolve their infrastructure to evade mitigation in Section~\ref{s:beyond}.

\section{Current Botnets \& Mitigations}\label{s:current} 

We first review the evolution of botnets and why we believe the next generation of botnets would subvert privacy infrastructures to evade detection and mitigation. Currently, bots are monitored and controlled by a botmaster, who issues commands. The transmission of theses commands, which are known as C\&C messages, can be centralized, peer-to-peer or hybrid~\cite{Bailey:2009}. In the centralized architecture the bots contact the C\&C servers to receive instructions from the botmaster. In this construction the message propagation speed and convergence is faster, compared to the other architectures. It is easy to implement, maintain and monitor. However, it is limited by a single point of failure. Such botnets can be disrupted by taking down or blocking access to the C\&C server. Many centralized botnets use IRC or HTTP as their communication channel. GT-Bots, Agobot/Phatbot~\cite{Liu:2009:BCA}, and clickbot.a~\cite{Daswani07theanatomy} are examples of such botnets. A significant amount of research focused on detecting and blocking them~\cite{gu2008botminer},~\cite{Cooke:2005},~\cite{Freiling:2005},~\cite{AbuRajab:2006},~\cite{John:2009},~\cite{Perdisci:2010}. To evade detection and mitigation, attackers developed more sophisticated techniques to dynamically change the C\&C servers, such as: Domain Generation Algorithm (DGA) and fast-fluxing (single flux, double flux).

Single-fluxing is a special case of fast-flux method. It maps multiple (hundreds or even thousands) IP addresses to a domain name. These IP addresses are registered and de-registered at rapid speed, therefore the name fast-flux. These IPs are mapped to particular domain names (e.g., DNS A records) with very short TTL values in a round robin fashion~\cite{fastflux}. Double-fluxing is an evolution of single-flux technique, it fluxes both IP addresses of the associated fully qualified domain names (FQDN) and the IP address of the responsible DNS servers (NS records). These DNS servers are then used to translate the FQDNs to their corresponding IP addresses. This technique provides an additional level of protection and redundancy~\cite{fastflux}. Domain Generation Algorithms (DGA), are the algorithms used to generate a list of domains for botnets to contact their C\&C. The large number of possible domain names makes it difficult for law enforcements to shut them down. Torpig~\cite{Stone-Gross:2009} and Conficker~\cite{Porras:2009} are famous examples of these botnets.

A significant amount of research focuses on the detection of malicious activities from the network perspective, since the traffic is not anonymized. For example~\cite{Antonakakis:DGA},~\cite{yadav2010detecting},~\cite{Yadav12},~\cite{Antonakakis:2011},~\cite{Antonakakis:2010},~\cite{exposure11} inspect the DNS traffic and use machine learning clustering and classification algorithms. BotFinder~\cite{botfinder:2012} uses the high-level properties of the bot's network traffic and employs machine learning to identify the key features of C\&C communications. DISCLOSURE~\cite{Disclosure:2012} uses features from NetFlow data (e.g., flow sizes, client access patterns, and temporal behavior) to distinguish C\&C channels. Other work~\cite{mohaisen2013babble},~\cite{west2014metadata} focus on endpoints' static metadata properties and the order of the high-level system events for threat classification.

The next step in the arms race between attackers and defenders was moving from a centralized scheme to a peer-to-peer C\&C. Storm~\cite{Holz:2008}, Nugache~\cite{nugache}, Walowdac~\cite{Stock:2009} and Gameover Zeus~\cite{GameoverZeus} are examples of such botnets. Some of these botnets use an already existing peer-to-peer protocol, while others use customized protocols. For example earlier versions of Storm used Overnet, and the new versions use a customized version of Overnet, called Stormnet~\cite{Holz:2008}. Meanwhile other botnets such as Walowdac and Gameover Zeus organize their communication channels in different layers.

Previous work studied specific mitigations against peer-to-peer botnets. For example BotGrep~\cite{BotGrep} uses the unique communication patterns in a botnet to localize its members by employing structured graph analysis. Zhang et al.~\cite{p2p:fingerprint} propose a technique to detect botnet P2P communication by fingerprinting the malicious and benign traffic. Yen and Reiter~\cite{FangReiter} use three features (peer churn, traffic volume and differences between human-driven and bot-driven behavior) in network flow to detect malicious activity. Coskun et al.~\cite{Coskun:2010} propose a method to detect the local members of an unstructured botnet by using the mutual contacts. As we can see, some of these techniques rely on observing the unencrypted traffic, therefore by using a privacy infrastructure such as Tor they can be evaded.

Very recently the use of Tor received more attention from malware and botnet authors. For example the new 64-bit Zeus employs Tor anonymity network in its botnet infrastructure~\cite{Zeus64}. It creates a Tor hidden service on the infected host and the C\&C can reach this infected hosts using their unique \texttt{.onion} address through Tor. Another example is ChewBacca~\cite{ChewBacca}, which uses Tor, and logs the keystrokes on the infected host and reports them back to the botmaster. The C\&C is an HTTP server that is hosted as a hidden service. Although using Tor and hidden services makes the detection and mitigation more difficult, these bots are still using the basic client-server model. This leaves them open to single point of failure.

\section{Privacy Infrastructure: Tor}\label{s:tor}

We envision OnionBots to rely on Tor for their operation. To better understand their potential and limitations, we briefly review the structure of Tor and hidden services. Tor~\cite{Tor} is the most widely used distributed low-latency anonymity-network. It helps users to resist censorship, and protects their personal privacy. Furthermore it allows users to hide their activities and location from government agencies and corporations. Clients establish anonymous communication by relaying their traffic through other Tor relays, called Onion Routers (OR). A client builds a circuit with the relays by negotiating symmetric keys with them. After building the circuit, the client sends the data in fixed sized \texttt{cells} and encrypts them in multiple layers, using the previously negotiated keys. Besides providing anonymous communication for clients, current implementation of Tor also offers anonymity for servers through hidden services.\\

The Tor hidden service architecture is composed of the following components:

\begin{figure}
\centering
\includegraphics[width=0.4\textwidth]{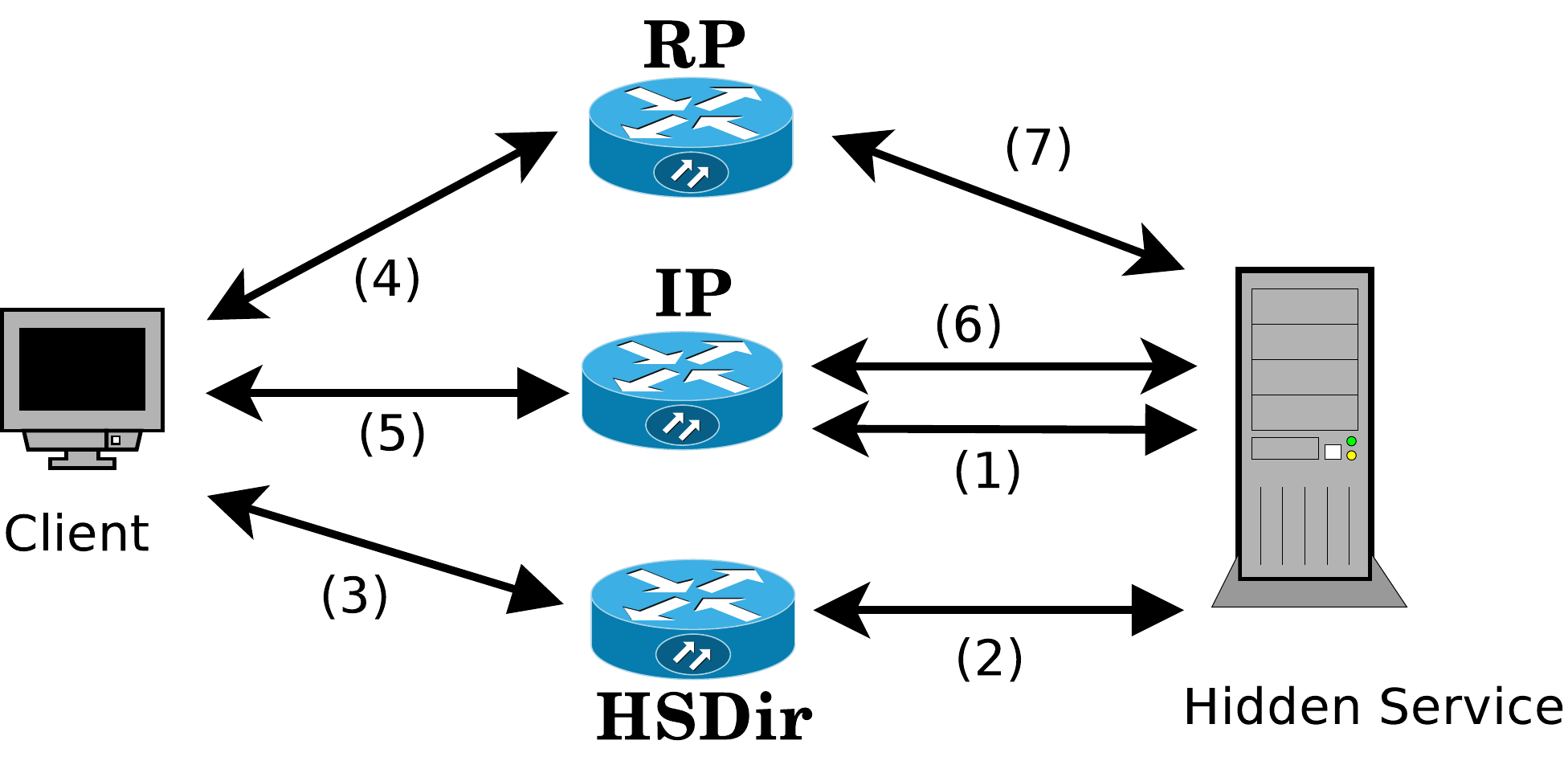}
\caption{Tor hidden service structure}
\label{f:TorHS}
\end{figure}

\begin{itemize}
\item{\emph{Server}, that runs a service (e.g., a web server).}
\item{\emph{Client}, that wishes to access the server.}
\item{\emph{Introduction Points (IP)}, a set of Tor relays, chosen by the hidden service, that forward the initial messages between the server and the client's Rendezvous Point.}
\item{\emph{Rendezvous Point (RP)}, a Tor relay randomly chosen by the client that forwards the data between the client and the hidden service.}
\item{\emph{Hidden Service Directories (HSDir)}, a set of Tor relays chosen by the server to store its descriptors.}
\end{itemize}

In order to make a service available via Tor, Bob (the service provider) generates an RSA key pair. The first 10 bytes of the SHA-1 digest of the generated RSA public key becomes the \texttt{Identifier} of the hidden service. The \texttt{.onion} hostname, is the base-32 encoding representation of the public key. As Figure~\ref{f:TorHS} illustrates the following steps take place to connect to a hidden service. Bob's Onion Proxy (OP) chooses a set of Tor relays to serve as his Introduction Points and establishes a circuit with them (step 1). After making the circuits, he computes two different service descriptors that determine which Tor relays should be chosen as its HSDirs (step 2). HSDirs are responsible for storing the hidden service descriptors, which change periodically every 24 hours and are chosen from the Tor relays that have the HSDir flag. This flag is obtained by relays that have been active for at least 25 hours. Later, in section~\ref{s:mitigation} we discuss the critical role that HSDirs can paly in mitigating OnionBots.\\

\noindent
\small \texttt{descriptor\mbox{-}id = H(Identifier~\textbar\textbar~secret-id-part)}\\
\small \texttt{secret\mbox{-}id\mbox{-}part = H(time-period~\textbar\textbar~descriptor-cookie \textbar\textbar~replica)}\\
\small \texttt{time\mbox{-}period = (current-time + permanent-id-byte * 86400 / 256) / 86400}\\
\normalsize

\begin{figure}
\centering
\includegraphics[width=0.45\textwidth]{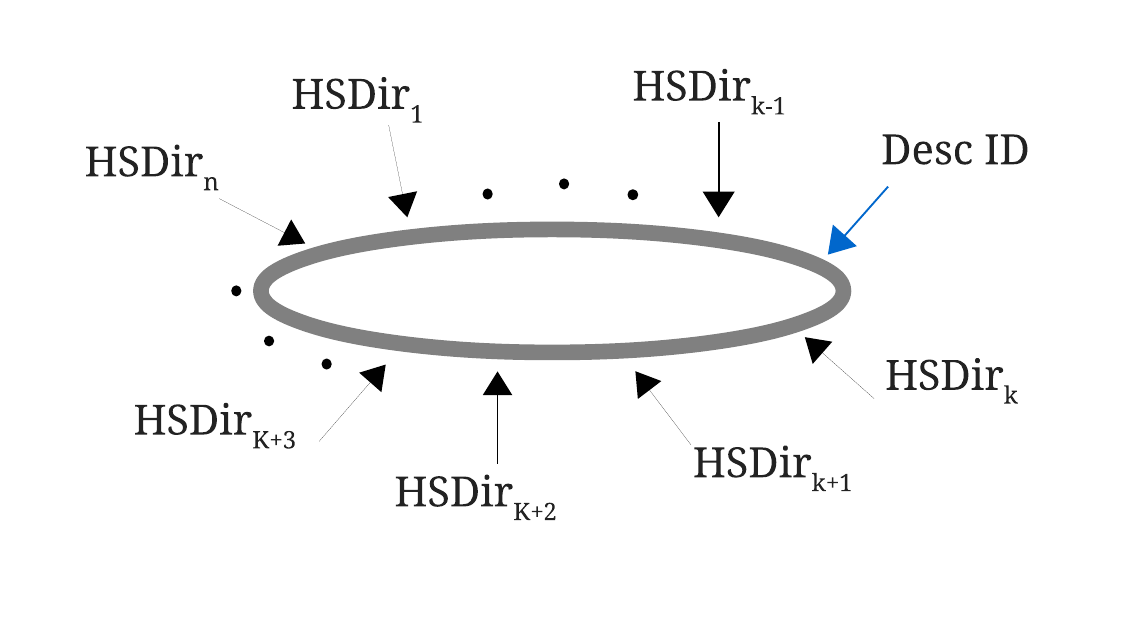}
\caption{Tor hidden service Directories (HSDir) Fingerprint}
\label{f:TorHSDir}
\end{figure}

\texttt{H} denotes the SHA-1 hash digest. \texttt{Identifier} is the 80 bit fingerprint (truncated SHA-1 digest of public key) of the hidden service. \texttt{Descriptor-cookie} is an optional 128 bit field. It can be used to provide authorization at the Tor network level and preventing unauthorized clients from accessing the hidden service. \texttt{Time-period} is used to periodically change the responsible HSDirs, and making the system more resilient. The \texttt{permanent-id-byte} prevents the descriptors from changing all at the same time. \texttt{Replica} takes values of 0 or 1. It is used to compute two different sets of descriptor IDs for a hidden service. The current implementation of Tor distributes each set of descriptor IDs in 3 different HSDirs, therefore for each hidden service there are a total of 6 responsible HSDirs. The list of Tor relays, which is called the \texttt{consensus} document, is published and updated every hour by the Tor authorities. If we consider the circle of the fingerprint of Tor relays as depicted in Figure~\ref{f:TorHSDir}, then if the descriptor ID of the hidden service falls between the fingerprint of HSDir$_{k-1}$ and HSDir$_{k}$, it will be stored on HSDir$_{k}$, HSDir$_{k+1}$ and HSDir$_{k+2}$.

When a client (Alice) wants to contact a hidden service, she first needs the hostname of the service.  Then from the \texttt{.onion} hostname, she computes the \texttt{Descriptor ID} of the hidden service and the list of its responsible HSDirs (step 3). To communicate with the hidden service, Alice first randomly chooses a Tor relay as her Rendezvous Point and makes a circuit to it (step 4). She then builds a new circuit to one of Bob's Introduction Points, and sends it a message. This message includes the Rendezvous Point's info and the hash of the public key of the hidden service (step 5). If the public key (sent by Alice) is recognized by the Introduction Point, it will forward the information to the hidden service OP (step 6). When Bob's OP receives the message it will extract the Rendezvous Point information and establishes a circuit with it (step 7).

This approach protects client's (Alice) IP address from Bob, and hides hidden service's (Bob) IP address from Alice, thus providing mutual anonymity for both the client and the server.

\section{OnionBot: a Cryptographic P2P Botnet}\label{s:onionbot} 
In this section, we look at the details of the proposed OnionBot, a novel non IP based construction that is immune to the current mitigation techniques. We explore different bootstrapping approaches and a distributed, self-healing, low-degree, low-diameter overlay peer-to-peer network formation.

\subsection{Overview}

OnionBot retains the life cycle of a typical peer-to-peer bot~\cite{narang2014peershark}. However, each stage has unique characteristics that make OnionBots different from current peer-to-peer botnets. As a result, existing solution are not applicable to them. For example, in the infection stage, each bot creates a \texttt{.onion} address and generates a key to indistinguishably encrypt the messages. In the rally stage, the bot dynamically peer with other bots that are the foundation of a self-healing network. Furthermore, while at the waiting stage, bots periodically change their address to avoid detection and mitigation. This new \texttt{.onion} addresses are generated from the key that is shared with the botmaster. This allows the botmaster to access and control any bot through the shared key, anytime, without revealing his identity.

\textbf{Infection}: is the first step in recruiting new bots, where each bots is also hardcoded with the public key of the botmaster. It can happen through traditional attack vectors such as phishing, spam, remote exploits, drive-by-download or zero-day vulnerabilities~\cite{verizon_report}. A great body of literature have focused on different spread mechanisms~\cite{Rossow13},~\cite{mobile:malware}. In this work we focus on the remaining stages of a bot's life cycle.

\textbf{Rally}: in order to join a botnet, the newly infected bots need to find already existing members of the network. In peer-to-peer network this process it called bootstrapping. For clarity reasons we use the same terminology in describing OnionBots. Based on the requirements of the network, the complexity and flexibility of bootstrapping techniques varies significantly. OnionBots necessitate a distributed mechanism to maintain a low-degree, low-diameter network. Such requirements, demands a bootstrapping mechanism that is able to evolve with the network. In section~\ref{ss:bootstrap} we discuss different techniques and their ramifications more in detail.

%On the other hand, OnionBots benefit from the decoupling of IP address and bots, which allows them to address such requirements. In section~\ref{ss:bootstrap} we discuss different techniques and their ramifications more in detail.

\textbf{Waiting}: in this state, a bot is waiting for commands from the botmaster. Generally the command transmissions can be pull-based (bots make periodic queries to the C\&C) or push-based (bot master sends the commands to the bots), and there are trade-offs in each mechanism. For example in the pull-based approach, if bots aggressively make queries for the C\&C messages, it allows faster propagation of commands. However it results in easier detection of C\&C and the bots. In the push-based approach, it is important to be able to reach to each bot, within reasonable steps. Furthermore, to prevent leakage of information about the botnet operation and topology, it should not be feasible for an adversary to distinguish the source, destination and the nature of the messages. Meanwhile, satisfying such requirements is not trivial in self-healing networks. Later in section~\ref{ss:cc} we discuss how in OnionBots, the bot master is able to access and control any bot.

\textbf{Execution}: at this stage the bots execute the commands given by the botmaster (e.g., bitcoin mining, sending spam~\cite{Thonnard:2011} or DDoS attack~\cite{cyber:attack}), after authenticating them. Recently botmasters started offering botnet-as-a-service~\cite{rent-service} as it was previously predicted by researchers in 2008~\cite{Hund:rambot}. Considering that the OnionBots make use of cryptographic block beyond the basic, trivial encryption/decryption of payload, it allows them to offer the botnet for rent. In section~\ref{ss:operation}, we explain how this can be done, in a distributed way, and without further intervention of the botmaster.

In the next sections we will focus on describing the key mechanisms of OnionBots.

\subsection{Bootstrap}\label{ss:bootstrap}

As mentioned previously the bootstrapping is an essential part of network formation in peer-to-peer networks. Additionally, in OnionBots, it provides the foundation for he self-healing graph construction. In the following, we study different approaches and their trade-offs. We discuss how these concepts should be adapted to the context of a privacy infrastructure such as Tor. Note that, the address of a peer list in our protocol refers to the \texttt{.onion} address of the peers, unless stated otherwise.

\begin{itemize}

\item \emph{Hardcoded peer list}: in this setting each bot has a list of other bots to contact at the beginning. Since the infections can be carried out by bots, this new peer lists can be updated. Each peer upon infecting another host will send a subset of its peer list. Each node in the original peer list will be included in the subset with probability \emph{p}. In the conventional botnets this scheme is vulnerable to detection and blacklisting. However, in OnionBots, the \texttt{.onion} address is decoupled form IP address, and changes periodically as it is described in section~\ref{ss:cc}. Therefore the current mitigations are not applicable.

\item \emph{Hotlists (webcache)}: this is conceptually similar to the hardcoded peer list. However each bot has a list of peers to query for other peers. In this setting, the adversary (defenders) will only have access to a subset of servers, since each bot only has a subset of the addresses, and these subsets can be updated upon infection or later in the waiting stage.

\item \emph{Random probing}: in this scheme a bot randomly queries the list of possible addresses, until it finds a bot listening on that address. Although it can be used in IPv4 and IPv6~\cite{Bless:2009} networks, it is not feasible in the context of Tor \texttt{.onion} addresses. Since the address space is intractable (to craft an address with specific first 8 letters, takes about 25 days~\cite{shallot}), to randomly query all possible \texttt{.onion} addresses requires probing an address space of size $32^{16}$.

\item \emph{Out-of-band communication}: the peer list can be transmitted through another infrastructure. For example by using a peer-to-peer network such as BitTorrent and Mainline DHT to store and retrieve peer lists, or by using social networks, such as Twitter, Facebook or YouTube.

\end{itemize}

We envision that OnionBots would use a customized approach based on hardcoded peer list and hotlists. As mentioned earlier in OnionBots the blacklisting of nodes is not practical, since their address changes periodically. In the following section we describe how OnionBots address the bootstrapping and recruitment during network formation and maintenance.

\subsection{Maintenance of the OnionBot Communication Graph}\label{ss:design}

\begin{figure*}
\centering
\includegraphics[width=0.9\textwidth]{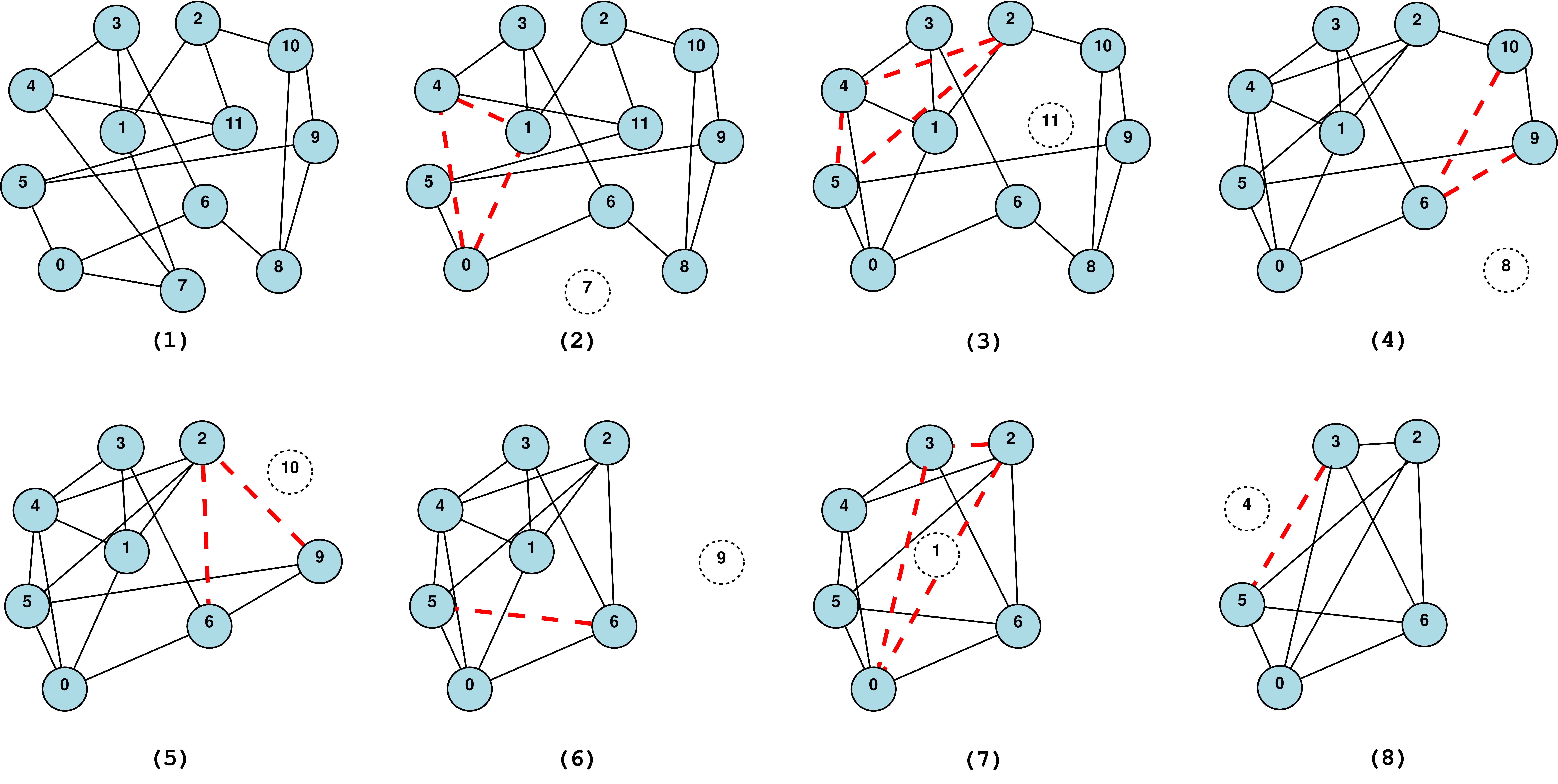}
\caption{Node removal and the self repairing process in a \emph{$3$-regular} graph with 12 nodes. The dashed red lines, indicate the newly established links between the nodes.}
\label{f:non-remove-node}
\end{figure*}

OnionBots form a peer-to-peer, self-healing network that maintains a low degree and a low diameter with other bots to relay messages. Peer-to-peer networks are broadly categorized as structured and unstructured~\cite{p2psurvey}, where both categories are used by botnets, and studied in previous work~\cite{nugache},~\cite{Holz:2008},~\cite{Starnberger:2008}. However, the already existing peer-to-peer networks are general in terms of their operations. Therefore, their design and resiliency is based on different assumptions and requirements. In the following, we propose a Dynamic Distributed Self Repairing (DDSR) graph, a new peer-to-peer construction that is simple, stealthy and resilient. Furthermore it is an overlay network, formed over a privacy infrastructure such as Tor.

\textbf{Neighbors of Neighbor Graph}: In this section, we introduce DDSR graph construct that is used in the network formation of OnionBots. The proposed construct is inspired by the knowledge of Neighbors-of-Neighbor (NoN). Previous work~\cite{non:2004} studied the NoN greedy routing in peer-to-peer networks, where it can diminish route-lengths, and is asymptotically optimal. In this work we discuss how NoN concepts can be used to create a self-healing network.

Consider graph $G$ with $n$ nodes ($V$), where each node $u_{i}$, $0 \leq i < n$, is connected to a set of nodes. The neighbors of $u_i$, are denoted as $N(u_{i})$. Furthermore, $u_i$ has the knowledge of nodes that are connected to $N(u_{i})$. Meaning that each node also knows the identity of its neighbor's neighbors. In the context of our work the identity is the \texttt{.onion} address.

\textbf{Repairing}: When a node $u_{i}$ is deleted, each pair of its neighbors $u_{j}, u_{k}$ will form an edge ($u_{j}$, $u_{k}$) iff ($u_{j}$, $u_{k}) \notin E$, where $E$ is the set of existing edges. Figure~\ref{f:non-remove-node} depicts the node removal and the self repairing process in a \emph{$3$-regular} graph with 12 nodes. The dashed red lines indicate the newly established links between the nodes. For example, as we can see if we remove one of the nodes (7), its neighbors (0, 1, 4) start to find a substitute for the delete node (7), to maintain the aforementioned requirements. In this case the following edges are created: (0, 1), (1, 4), and (1, 4).

The basic DDSR graph outlined in the previous paragraph does not deal with the growth in the connectivity degree of each node, denoted by $d(u)$; after multiple deletions the degree of some nodes can increase significantly. Such increase in the nodes degree is not desirable for the resiliency and stealth of the botnet, therefore we introduce a pruning mechanism to keep the nodes degree in the range $[d_{min}, d_{max}]$. Note that $d_{min}$ is only applicable as long as there are enough surviving nodes in the network. \\

\textbf{Pruning}: Consider the graph $G$, when a node $u_{i}$ is deleted, each one of its neighbors, starts the repairing process. However this scheme causes degree of the neighbors of node $u_{i}$, to increase significantly after $t$ steps (deletions). To maintain the degree in the aforementioned range($[d_{min}, d_{max}]$), each neighboring node of the deleted node ($u_{i}$), deletes the highest degree node from its peer list. If there is more than one such candidate, it randomly selects one among those for deletion, until its degree is in the desired range. Removing the nodes with the highest degree, maintains the reachability of all nodes, and the connectivity of the graph.

\textbf{Forgetting}: In the proposed OnionBot, nodes forget the \texttt{.onion} address of the pruned nodes. Additionally, to avoid discovery, mapping and further blocking, each bot can periodically change his \texttt{.onion} address and announce the new address to his current peer list. The new \texttt{.onion} address is generated based on a secret key and time. This periodic change is possible because of the decoupling between IP address and the bots that is provided by Tor. Later, in section~\ref{ss:cc} we explain how the C\&C is able to directly reach each bot, even after they change their address.

\subsection{Command and Control Communication}\label{ss:cc}

In this section, we show how Tor enables a communication channel with C\&C that is stealthy, and resilient. As mentioned before, OnionBot is a non IP address based construction, therefore current techniques to detect and disable the C\&C (e.g., IP blacklisting and DNS traffic monitoring) are ineffective.

In OnionBot we assume two classes of messages: 1) messages from C\&C to the bots and 2) messages from bots to C\&C. The messages from C\&C can be either directed to an individual node(s) (e.g., a maintenance message telling a bot to change it peers) or directed to all bots (e.g., DDoS attack on example.com). Furthermore, the botmaster can setup group keys to send encrypted messages for a group of bots. While a bot can tell the difference between a broadcast message and messages directed individual bot(s), it is not able to identify the source, the destination and the nature of these messages. Therefore the authorities are not able to detect different messages and drop harmful message and only allow the maintenance message to pass through. As a result they can not create the illusion that the botnet is operational, when it is actually neutralized.

In OnionBot, the bots report their address to C\&C, and establish a unique key to be shared with botmaster ats the infection/rally stage. This allows C\&C to have direct access to the bots, even after they change their \texttt{.onion} address. Each bot generates a symmetric key, $K_B$ and reports it to the C\&C; by encrypting $K_B$ with the hard coded public key of the C\&C ($\{ K_B\}_{PK_{CC}}$). After establishing the key, bots can periodically change their \texttt{.onion} address based on a new private key generated using the following recipe, \texttt{generateKey($PK_{CC}$, H ($K_B$, $i_{p}$))}. Where, \texttt{H} is a hash function, and $i_{p}$ is the index of period (e.g., day). All messages are of the same fixed size, as they are in Tor. Furthermore, to achieve indistinguishability between all messages, we use constructions such as Elligator~\cite{Bernstein:2013}. As a result  no information is leaked to the relaying bots.

\subsection{Operation}\label{ss:operation}

While many current botnets lack adequate secure communications~\cite{Rossow:2013:SPM} (e.g., sending messages in plaintext) that leaves them open to hijacking, the OnionBot's communication is completely encrypted since it uses Tor and SSL. Note that the the encryption keys are unique to each link. Table~\ref{t:botnet-crypto} summarizes a number of botnet families and their lack of adequate crypto blocks, after discovery and reverse engineering~\cite{Rossow:2013:SPM}. Furthermore, we introduce new cryptographic blocks that enable the OnionBot to offer new services, such as a botnet-for-rent~\cite{Hund:rambot} and a distributed computation platform for rent.

\begin{table}
\centering
\scalebox{1.3}{
\begin{tabular}{ | c | c | c | c | }\hline

	\textbf{Botnet} & \textbf{Crypto} & \textbf{Signing} & \textbf{Replay} \\ \hline
	Miner & none & none  &  yes	\\ \hline
	Storm & XOR & none  &  yes	\\ \hline
	ZeroAccess v1 & RC4 & RSA 512 & yes \\ \hline
	Zeus & chained XOR & RSA 2048 & yes \\ \hline

\end{tabular}
}
\caption{Cryptographic use in different botnets.}
\label{t:botnet-crypto}
\end{table}

To achieve the aforementioned services, we need to account for three aspects of the messages: 1) authenticity, 2) expiration time, and 3) legitimacy. Public key encryption and certificates based on the chain of trust, are suitable candidates to solve the authenticity and the legitimacy of the messages, and the expiration of the message (rental contract term) can be addressed by using timestamps. In the following we describe the details of such operation.

Imagine Trudy wishes to rent the botnet from Mallory, and every bot has the public key of Mallory hardcoded. Trudy sends her public key $PK_{T}$ to Mallory, to be signed by the private key of Mallory $SK_{M}$. The signed message ($T_{T}$) acts as a token containing $PK_{T}$, an expiration time, and a list of whitelisted commands. When Trudy wants to issue a command to her rented bots, she signs her command by using her private key $SK_{T}$ and includes $T_{T}$. The bots are able to verify the legitimacy of such commands, by looking at the token and the signature of the message.

As a bussiness operation, Trudy pays Mallory using Bitcoin, where the whole transaction can be carried out over Silk Road 2.0~\cite{silk2}. Furthermore Mallory can instruct her bots to install computation platforms such as Java Virtual Machine (JVM). By doing so, she can also offer a cross-platform distributed computation infrastructure to carry out CPU intensive tasks, such as bitcoin mining or password cracking.

\section{OnionBots Evaluation}\label{s:eval}

\begin{figure*}
     \centering
     \subfloat[][without pruning]{\includegraphics[scale=0.24]{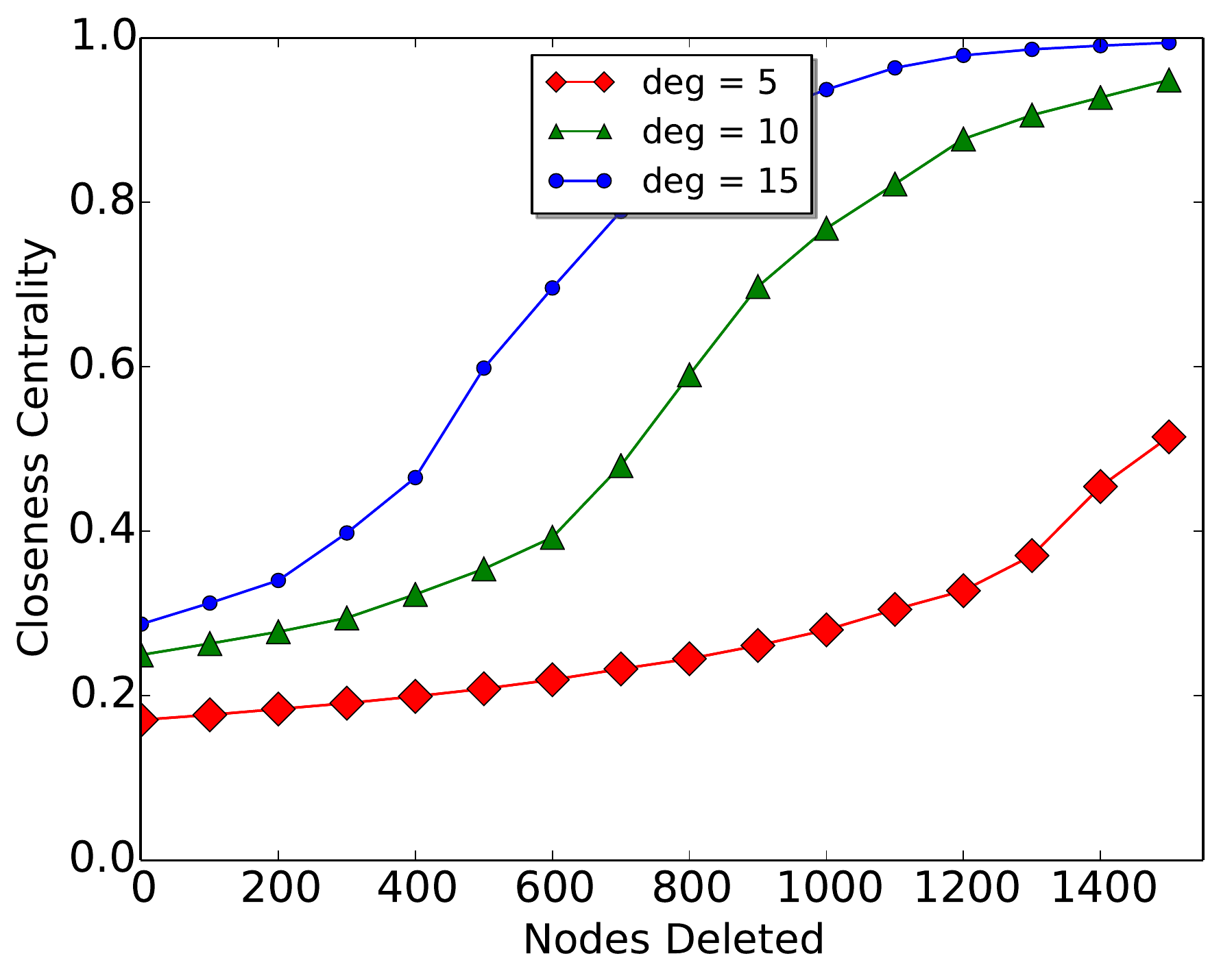}\label{f:closeness-wo-prune}}
     \subfloat[][with pruning]{\includegraphics[scale=0.24]{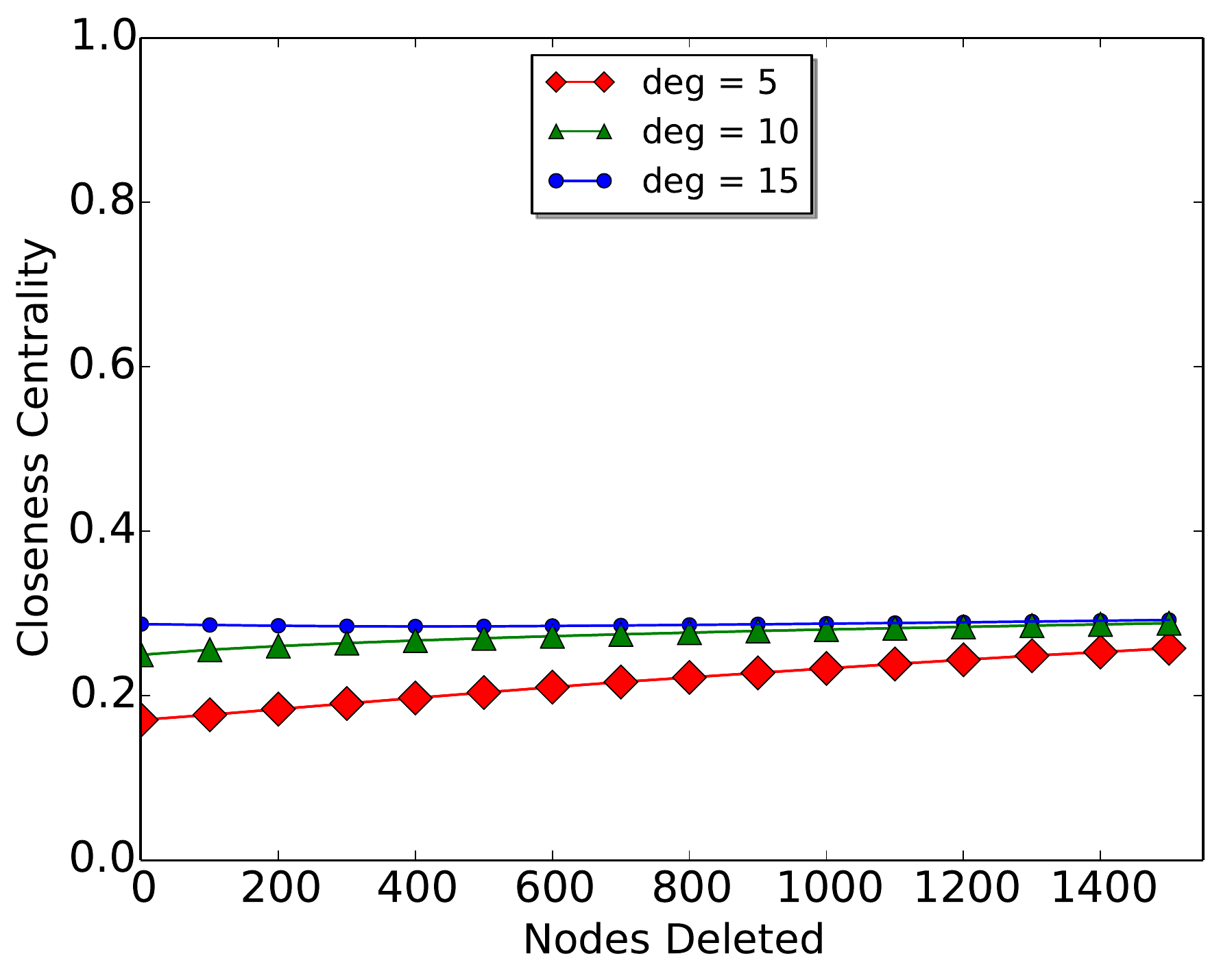}\label{f:closeness-w-prune}}
	 \subfloat[][without pruning]{\includegraphics[scale=0.24]{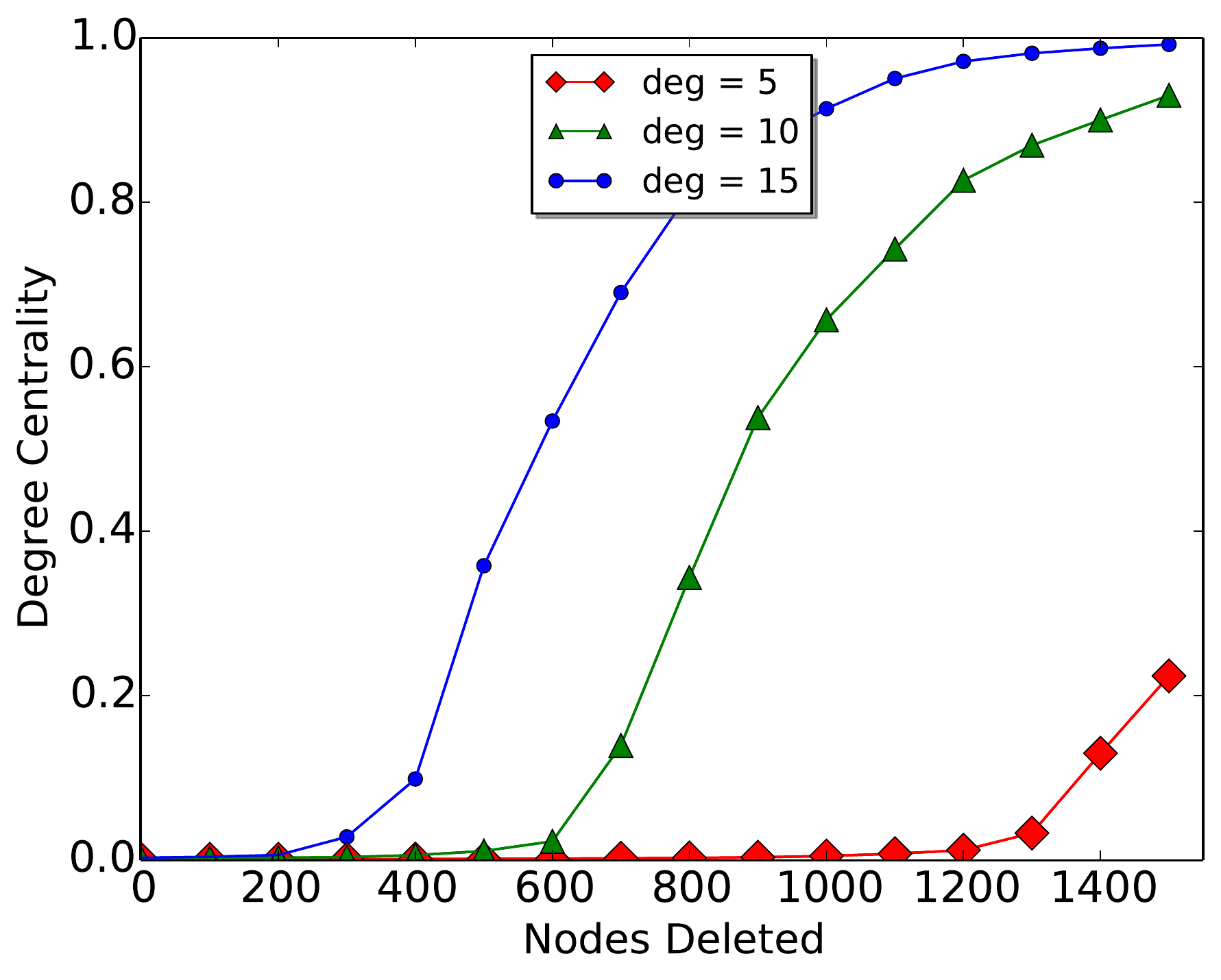}\label{f:deg-cent-wo-prune}}
     \subfloat[][with pruning]{\includegraphics[scale=0.24]{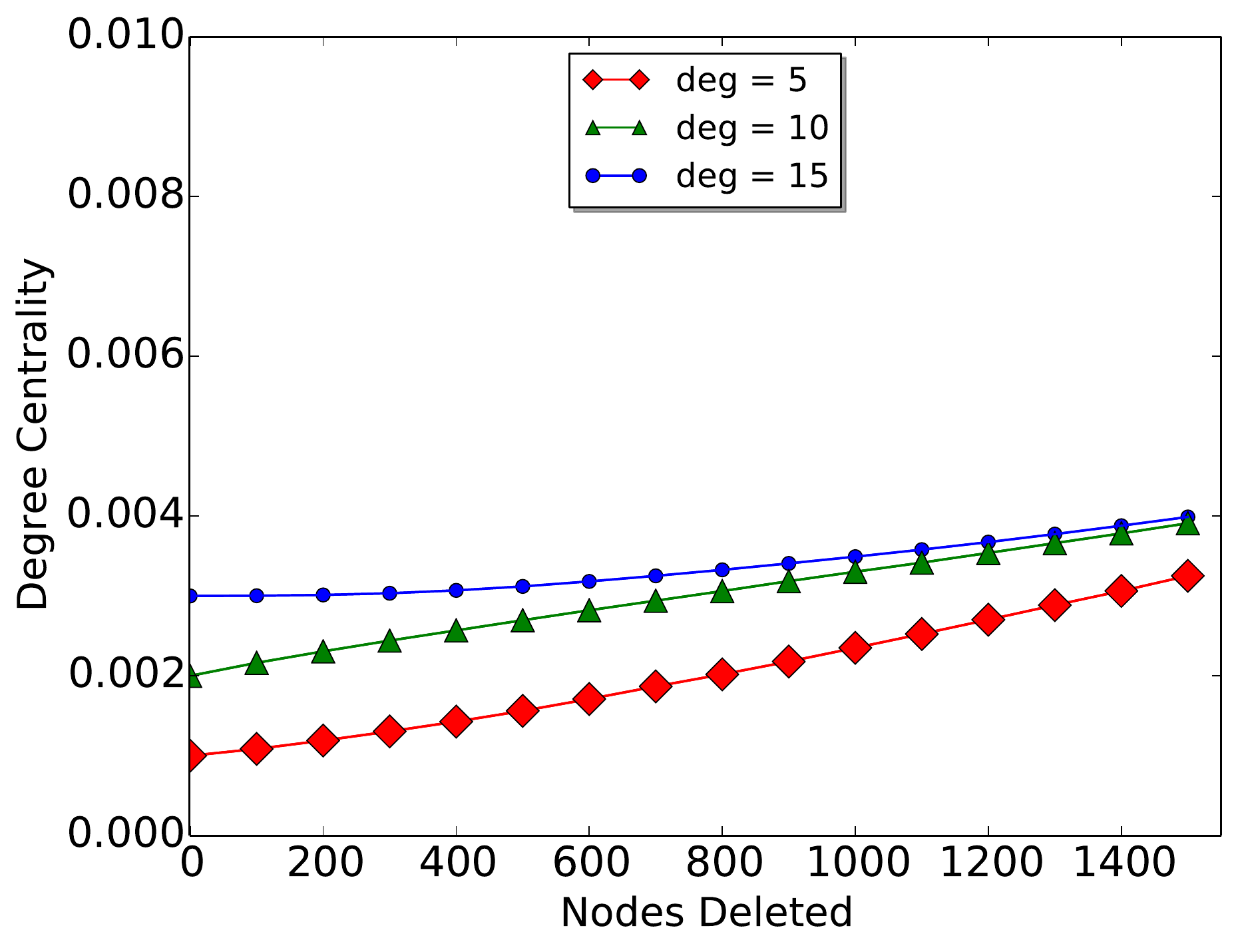}\label{f:deg-cent-w-prune}}
     \caption{The average closeness centrality (\ref{f:closeness-wo-prune},~\ref{f:closeness-w-prune}) and degree centrality (\ref{f:deg-cent-wo-prune},~\ref{f:deg-cent-w-prune}) of nodes in a \emph{k-regular} graph, ($k = 5, 10, 15$) with 5000 nodes after 30\% node deletions, with and without pruning.}
     \label{f:closeness}
\end{figure*}

To evaluate the envisioned OnionBots, we look at two aspects: the self-healing network formation resilience performance and the resilience to analysis techniques, such as botnet mapping, hijacking, or even assessing the size of the botnet. The NoN look-ahead routing is proven to be asymptotically optimal, however such claims have not been studied in the context of self-healing networks. Although it is desirable to rigorously prove properties of such networks, in this work we use empirical data and simulations for evaluation.

\subsection{Mapping OnionBot}

OnionBots provide a more resilient structure by using features available in the Tor network that previous botnets lack. All OnionBot nodes are directly accessible, even those running behind NAT, compared to previous work~\cite{WangDSN}. If a bot is captured and the address (\texttt{.onion}) of other bots is revealed, it is still not practical to block the bots. Additionally, bots can periodically change their \texttt{.onion} address, and share it only with their operational peers. Therefore limiting the exposure of the bot's address. As a result one single \texttt{.onion} could be blocked as we later discuss in section~\ref{s:mitigation}, but it is not feasible to block all of them.

\subsection{OnionBot Network Resilience}

To evaluate the resiliency and performance of the self repairing construction we use some of the metrics that are used in graph theory, such as the changes on graph centrality after node deletions. Centrality metrics examined in previous studies~\cite{Yen:2012},~\cite{ideal:infra} include closeness centrality and degree centrality.\\

The \emph{closeness centrality} of node $u$, is the inverse of sum of the shortest paths between node $u$ to all $n-1$ other nodes. Since the sum of distances depends on the number of nodes, it is normalized.

\begin{equation*}
	 C(u) =  {n -1 \over{ \sum_{v \neq u} d(u,v)}}
\end{equation*}

Where n is the number of nodes, and $d(u,v)$ is the shortest path between nodes $u$ and $v$. It is an indication of how fast messages can propagate in the network, from a node, $v$, to all other nodes sequentially.

The \emph{degree centrality} of a node $u$ is the number of its neighbors. The degree centrality values are normalized by the maximum possible degree in graph $G$. Therefore it is the fraction of nodes that $u$ is connected to. It is an indication of immediate chance of receiving whatever is flowing through the network (e.g., messages).

Another metric that we explore and is overlooked by previous work, is the \emph{diameter} of the graph. The diameter of a graph is defined as the longest shortest path (geodesic distance) in the graph. It is formally defined as the maximum of $d(u,v)$, $\forall u,v$, and provides a lower bound on worst case delay.

While a theoretical analysis is more desirable, it is also much harder. In the following, we resort to simulation to get a good sense of the properties of OnionBot. We simulate the node deletion process in a \emph{k-regular} graph, ($k = 5, 10, 15$) of 5000 nodes, with up to 30\% (1500) node deletions. Figure~\ref{f:closeness} illustrates the average closeness centrality with pruning (Figure~\ref{f:closeness-w-prune}) and without pruning (Figure~\ref{f:closeness-wo-prune}). As we can see in Figure~\ref{f:closeness-wo-prune}, closeness centrality of the nodes is stable, and even after node deletion, it does not decrease. Furthermore we measure the degree centrality of the nodes in the aforementioned graph, with pruning (Figure~\ref{f:deg-cent-w-prune}) and without pruning (Figure~\ref{f:deg-cent-wo-prune}). As we can see, the degree of nodes increases significantly after node deletions without pruning. Low degree centrality is desirable in advanced persistent attacks (APT). It decreases the chances of detection and take down, because of maintaining a low profile and avoiding to raise the alarm. For example Stuxnet only infected maximum of three other nodes~\cite{stuxnet}, to slow down its spread and avoid detection.

\begin{figure}
     \centering
     \subfloat[][]{\includegraphics[scale=0.24]{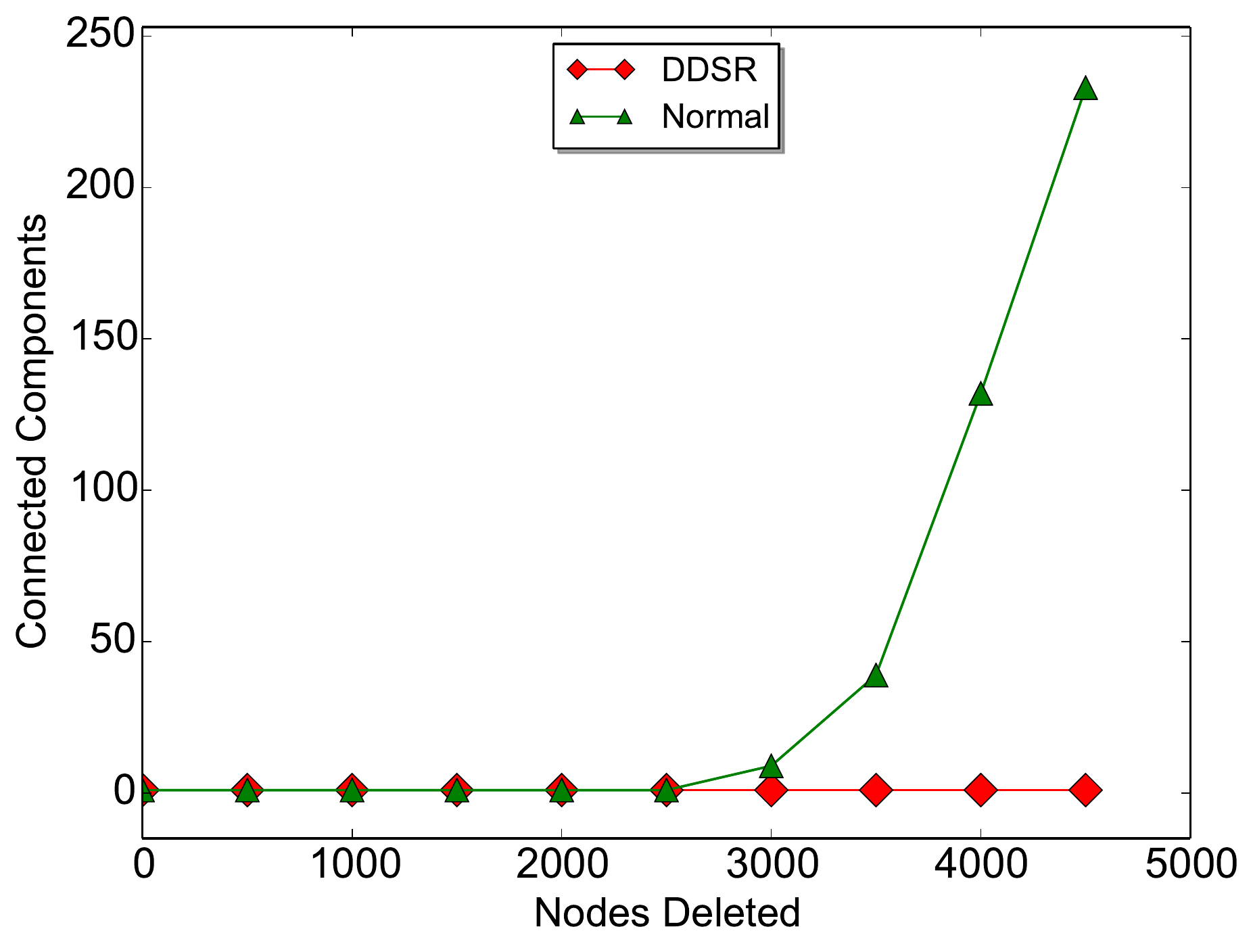}\label{f:5000_conn_comp}}
     \subfloat[][]{\includegraphics[scale=0.24]{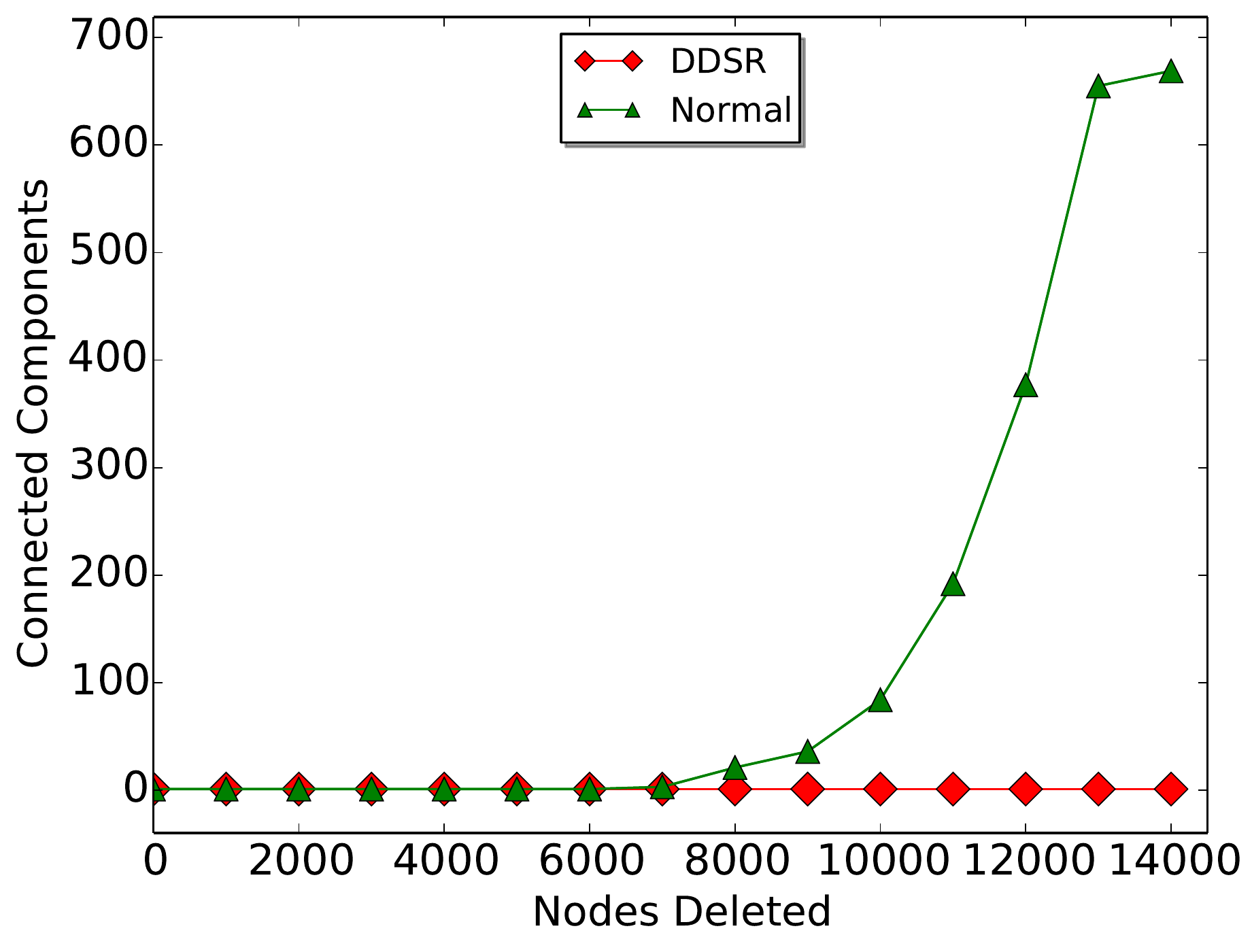}\label{f:15000_conn_comp}}\\
     \subfloat[][]{\includegraphics[scale=0.24]{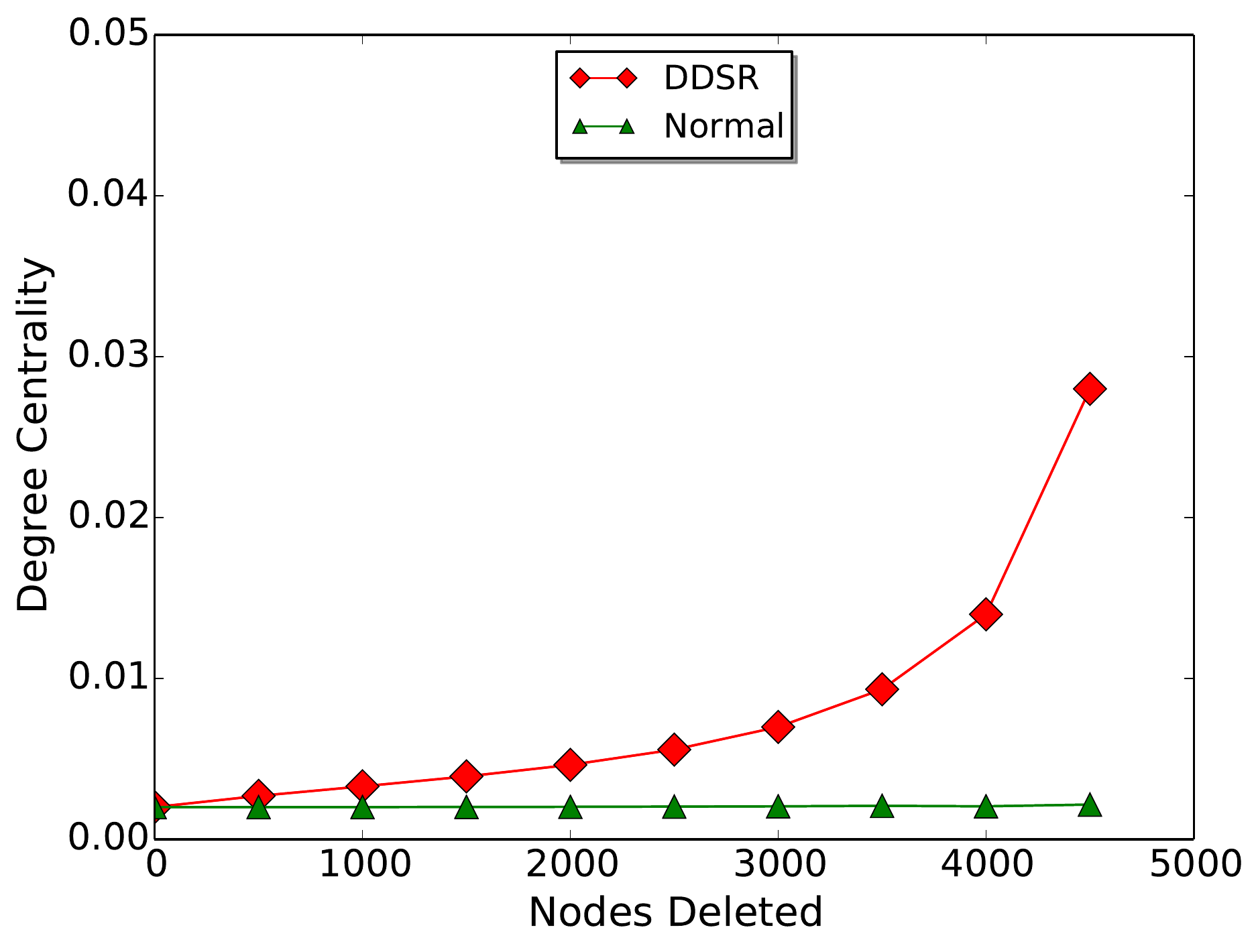}\label{f:5000_deg_cent}}
     \subfloat[][]{\includegraphics[scale=0.24]{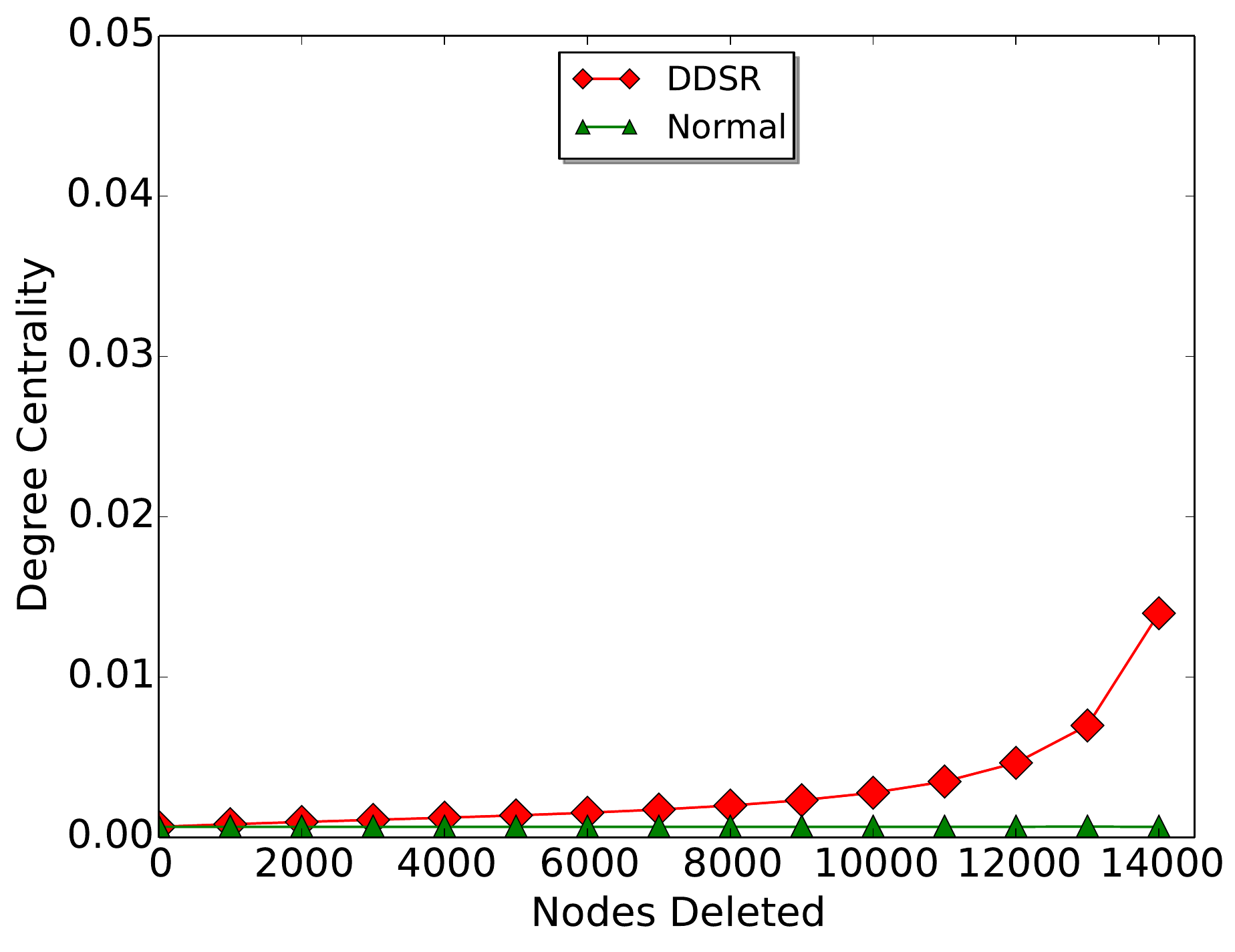}\label{f:15000_deg_cent}}\\
     \subfloat[][]{\includegraphics[scale=0.24]{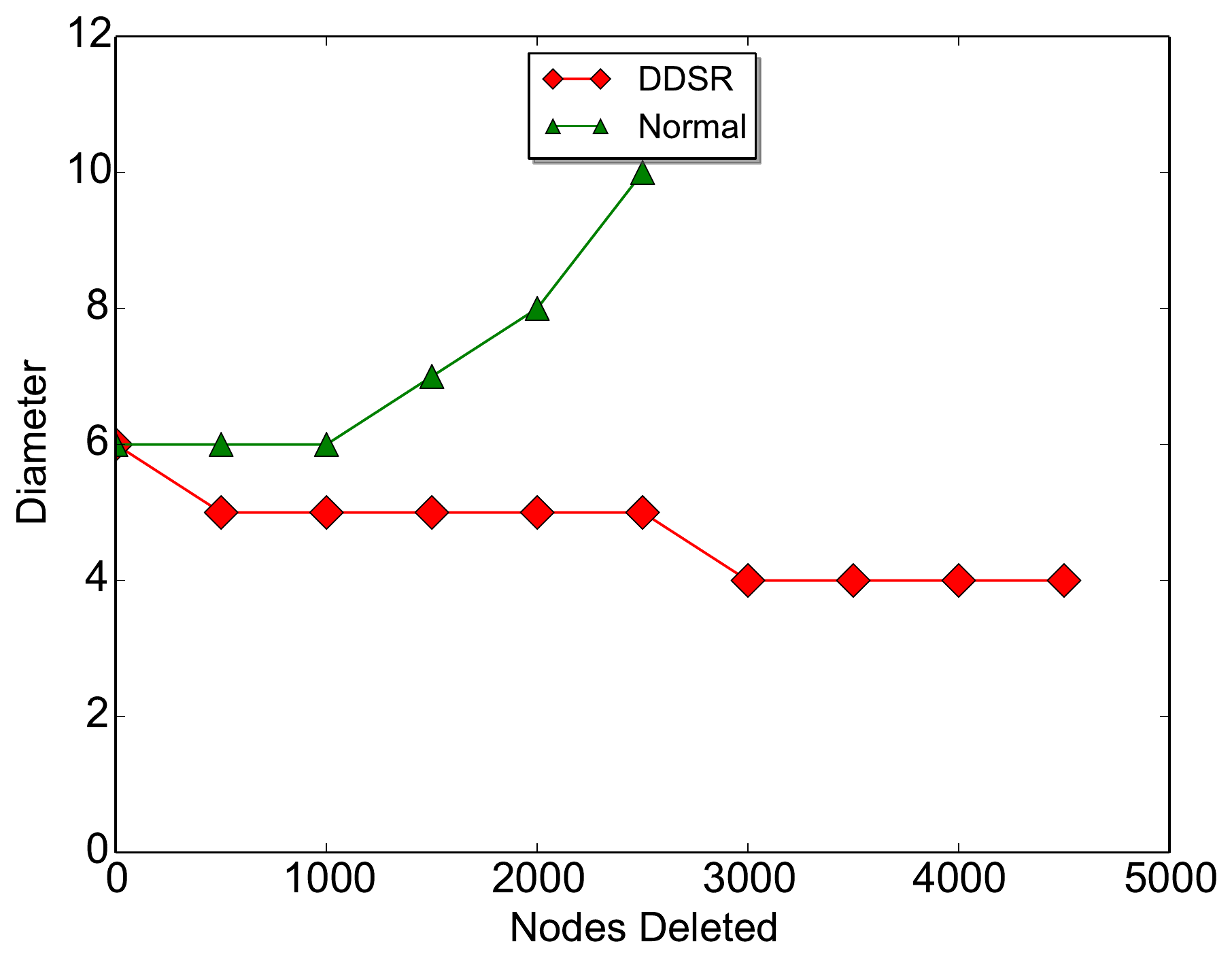}\label{f:5000_diameter}}
     \subfloat[][]{\includegraphics[scale=0.24]{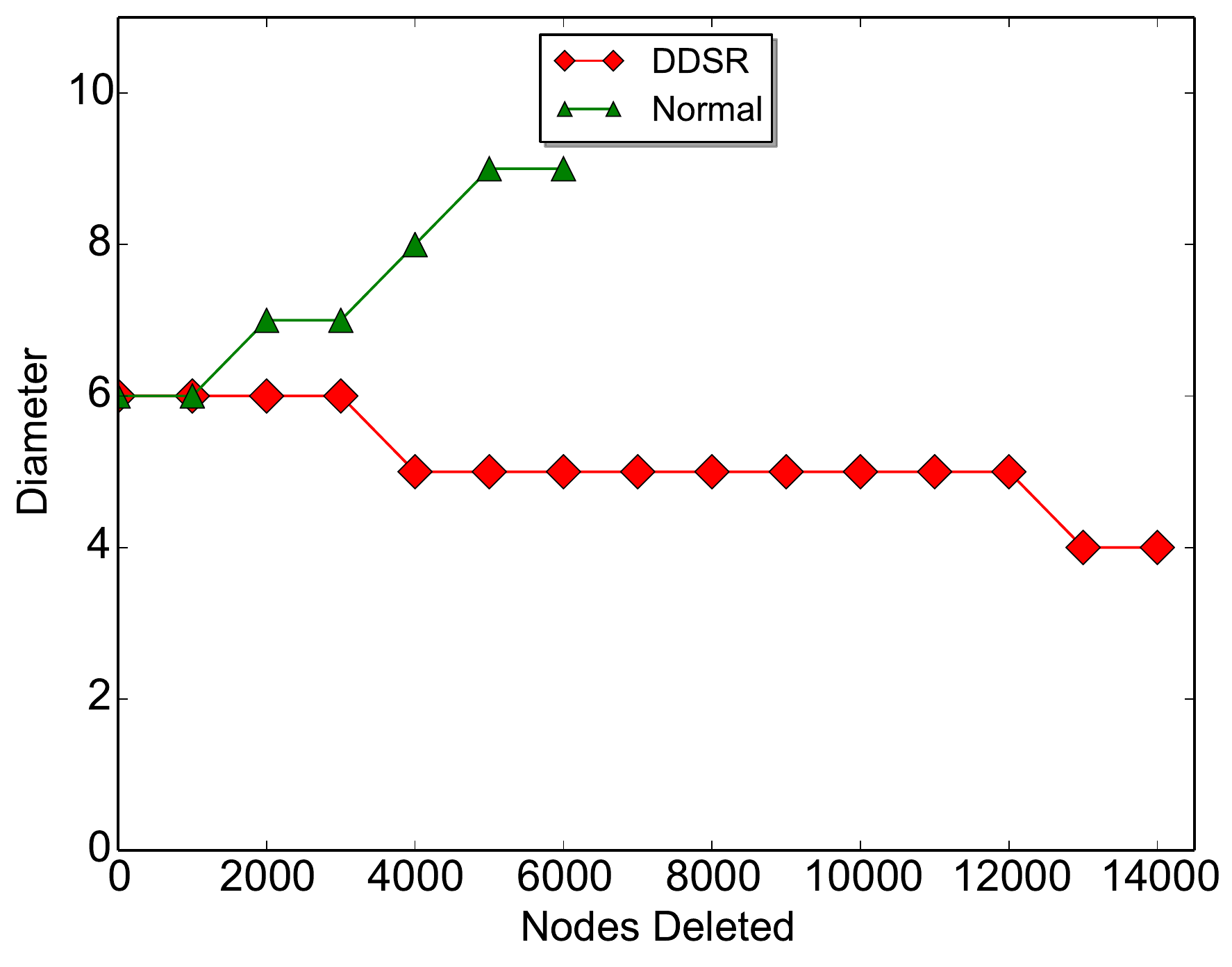}\label{f:15000_diameter}}
     \caption{Graphs depicting the number of connected components, average degree centrality and graph diameter, after incremental node deletions, in a \emph{$10$-regular} graph of 5000 (left side) and 15000 (right side) nodes.}
     \label{f:multi-nodes}
\end{figure}

To better understand the effect of size, we simulate a small botnet of size 5000~\cite{Yen:2012} and a medium botnet of size 15000. Figure~\ref{f:multi-nodes} depicts the aforementioned metrics. As we can see in Figures~\ref{f:5000_conn_comp} and~\ref{f:15000_conn_comp}, the self-repairing graph remains connected even when a large portion (90\%-95\%) of the nodes are deleted, compared to a normal graph (a graph with no self-repairing mechanism). Note that, in a normal graph after 60\% node deletion, the number of partitions increases sharply. As we can see in Figures~\ref{f:5000_deg_cent} and~\ref{f:15000_deg_cent}, the degree centrality slightly increases in the DDSR compared to a normal graph, since the healing process ensures that the degree of the nodes stays within the range. However, as we remove the nodes in a normal graph, the diameter increases until the graph is partitioned, where the diameter is infinite. In OnionBot, as the nodes are deleted and the number of nodes decreases, the diameter of the graph also decreases accordingly (Figures~\ref{f:5000_diameter} and~\ref{f:15000_diameter}).

\section{Mitigation of Basic OnionBots}\label{s:mitigation} 

\begin{figure}
\centering
\includegraphics[width=0.4\textwidth]{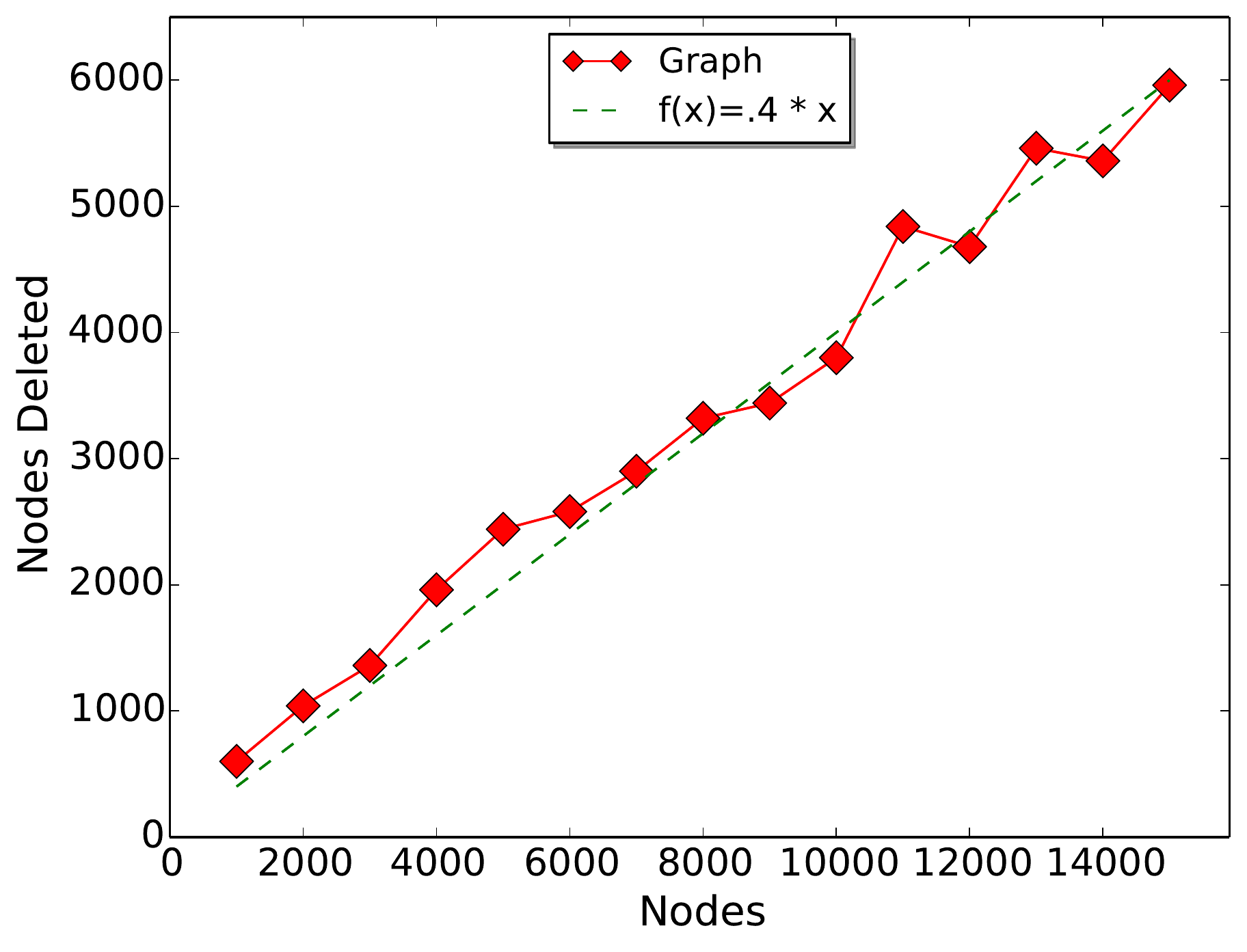}
\caption{Partitioning of graphs after removing nodes. The network becomes partitioned after removing on average about 40\% of the nodes, in \emph{$10$-regular} graphs of size n=1000, to n=15000.}
\label{f:graph-partitioned}
\end{figure}

\begin{figure*}
\centering
\includegraphics[width=1\textwidth]{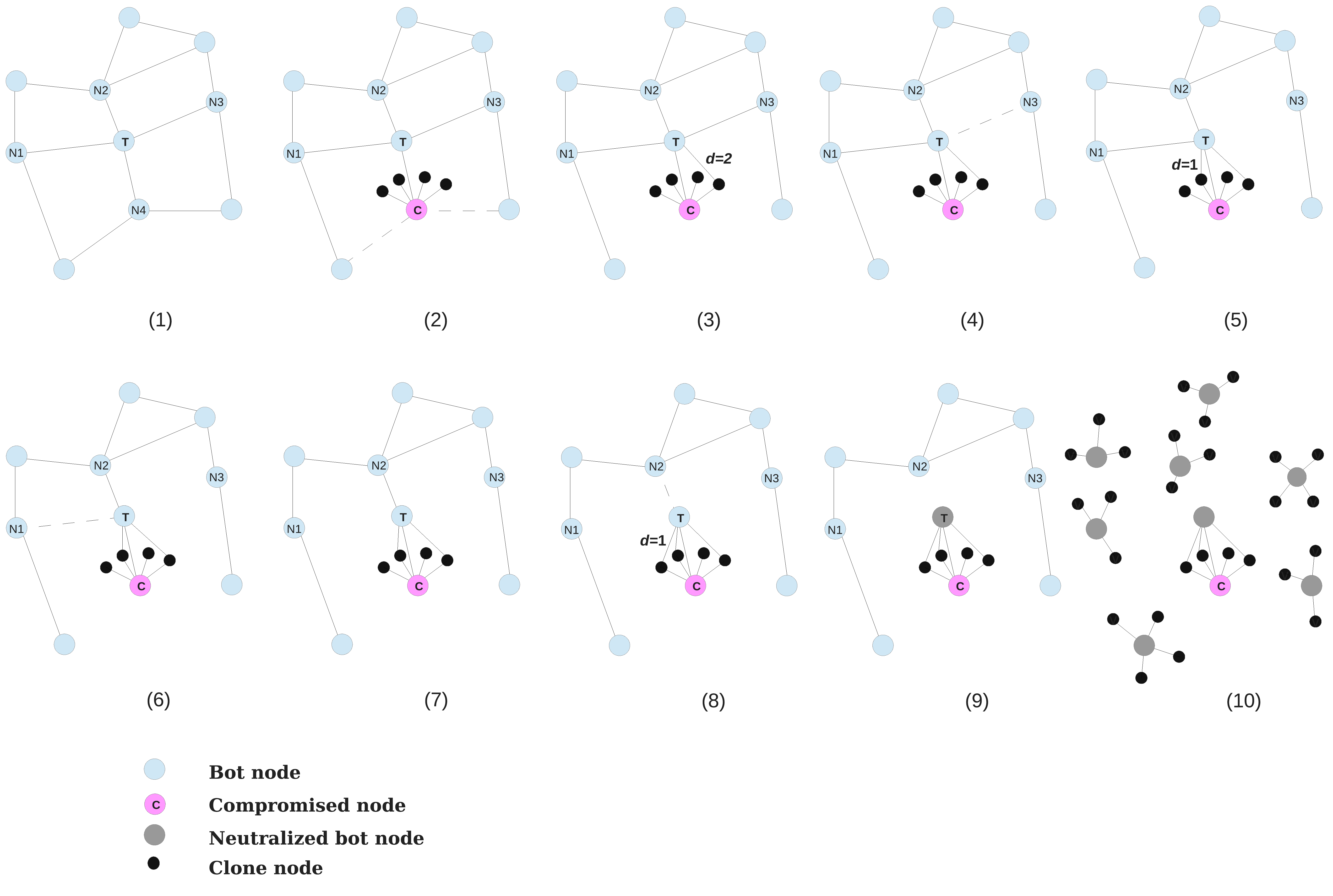}
\caption{Soaping attack: node T is under attack by the compromised node C and its clones. In each step one of the clones initiates the peering process with the T, until it is contained. After several iterations, the network is partitioned, and the botnet is neutralized.}
\label{f:soap-node}
\end{figure*}

In this section we look at different mitigation strategies against OnionBots. Mitigation and detection can take place at different levels, such as host level or network level. Host level remediation techniques include the use of anti-virus software and frameworks such as the Security Behavior Observatory (SBO)~\cite{forget2014security}. Because of the scaling limitation of such techniques, and the fact that compromised hosts are rarely updated or patched, we focus on the network level strategies.

Many of the current detection and mitigation mechanisms are IP-based (e.g., the WOMBAT attack attribution method~\cite{Dacier:2009}), and use the network traffic patterns or DNS queries to distinguish legitimate traffic from malicious traffic. However current solutions do not work with the OnionBots, since the Tor traffic is encrypted, non IP-based, and there are no conventional DNS queries to resolve the \texttt{.onion} addresses. Furthermore, even if an adversary captures a bot sample (e.g., by using honeypots or other similar techniques), and recovers the \texttt{.onion} address of its peers, he is still unable to directly map these addressees to their corresponding IP addresses and take down the infected hosts. Since the proposed construction offers a self-repairing low-degree, low-diameter network even after taking over a large portion of the bots, the botnet remains functional. As Figure~\ref{f:graph-partitioned} shows, an adversary needs to take down about the 40\% of the bots simultaneously, to even partition the network into 2 subgraphs. Note that, it means there is not enough time for the graph to self-repair. As we can see the conventional solutions that ignore the privacy infrastructure construction of OnionBots are not effective, therefore we need to adapt our detection and mitigation methods, and integrate them into the foundation of such infrastructures. In this section we divide network level mitigations into two categories; techniques that are generic to Tor, and schemes that are specific to OnionBot. In particular, we propose a new OnionBot specific mitigation method, called Sybil Onion Attack Protocol (SOAP).

\subsection{Targeting OnionBots Through Privacy Infrastructures}\label{sss:tor-mitig}

Generic mitigations targeting Tor are based on denying access to the bots through the \texttt{HSDirs}. As described before, the list of \texttt{HSDirs} can be calculated by any entity who knows the \texttt{.onion} address (in case there is no \texttt{descriptor-cookie}). Hence an adversary can inject her relay into the Tor network such that it becomes the relay responsible for storing the bot's descriptors. Since the fingerprint of relays is calculated from their public keys, this translates into finding the right public key~\cite{trawltor}. Nevertheless, it should be noted that an adversary needs to position herself at the right position in the ring at least 25 hours before (it takes 25 hours to get the \texttt{HSDir} flag). It is difficult to mitigate against many bots, since the adversary requires the computation power, and a prior knowledge of the \texttt{.onion} addresses. Furthermore it disrupts the operation and user experience of the Tor network.

A more long term approach involves making changes to the Tor, such as use of CAPTACHAs, throttling entry guards and reusing failed partial circuits, as described in~\cite{hopper2014}. Having said that, these mitigations are limited in their preventive power, open the door to censorship, degrade Tor's user experience, and are not effective against advanced botnets such as OnionBot.

\subsection{Sybil Onion Attack Protocol (SOAP)}\label{sss:soap}

We devised a mitigation mechanism that uses OnionBots' very own capabilities (e.g., the decoupling of IP address and the host) against them. We first overview the attack here, and then provide a step by step explanation as depicted in Figure~\ref{f:soap-node}. To attack the botnet and neutralize it, we first need to find the bots \texttt{.onion} address. This can be done, either by detecting and reverse engineering an already infected host, or by using a set of honeypots. Although this is not a trivial task, and requires a significant amount of effort, it allows us to infiltrate the botnet, traverse its network, and identify the other bots. After identify the bots \texttt{.onion} address, we run many hidden services, disclosing a subset of these as neighbors to each peer we encounter, so gradually over time our clone nodes dominate the neighborhood of each bot and contain it. Note that we can run all of the clones on the same machine because of the decoupling between the IP address and the host.

Figure~\ref{f:soap-node} depicts the soaping attack in different steps. Node $T$ is the target of the soaping attack, nodes $N_i$, are its neighboring bot nodes, and nodes $C$ are the adversary (e.g., the authorities), and his clones, which are represented with small black circles. In step 1, the botnet is operating normally, and none of $T$'s neighbors are compromised. In step 2, one of its peers, $N_4$, is compromised. Then, $N_4$ (now depicted as $C$), makes a set of clones (the small black circles). In step 3, a subset of $C$'s clones, start the peering process with $T$, and declare their degree to be a small random number, which changes to avoid detection (e.g., d=2). Doing so increases the chances of being accepted as a new peer, and replacing an existing peer of $T$. In step 4, $T$ forgets about one of its neighbors with the highest degree, $N_3$, and peers with one the clones. The clones repeat this process until $T$ has no more benign neighbors (steps 5-8). As a result, $T$ is surrounded by clones and is contained (step 9). As we can see after many iterations the adversary can partition the network into a set of contained nodes, and neutralizing the botnet. In section~\ref{s:beyond}, we discuss how the attackers can withstand the soaping attack, by using sophisticated detection and mitigation mechanisms, such as probing and proof of work.

\section{Beyond Basic OnionBots}\label{s:beyond}

As the authorities design mitigation techniques against advanced botnets such as OnionBost, the attackers will evolve their infrastructure to evade detection and blocking. This arms race, results in the design and deployment of more sophisticated generation of OnionBots that limit the mitigation efforts. Such increased complexity, introduces new challenges, both for authorities and attacker. In the following we briefly discuss a few of them.

\subsection{Overcoming SOAP}\label{ss:basicsoapmitig}

Although basic OnionBots are susceptible to the soaping attack, the attacker can mitigate against them, by using proof of work and rate limiting. In the proof of work scheme each new node needs to do some work before being accepted as a peer of an already existing node. As more nodes request peering with a node, the complexity of the task is increased to give preference to the older nodes. The same approach can be used in the rate limiting, where the delay of accepting new nodes is increased proportional to the size of peer list. Although such actions increase the adversarial resilience of the network, they also decrease the flexibility and the recoverability of the network. Finding the right balance between the recoverability and adversarial resilience is an open question that the research community needs to study in detail to allow the preemptive design of effective detection and mitigation mechanisms. Additionally the attacker can make the NoN construction more resilient, and create a \emph{SuperOnionBot}, by fully utilizing the decoupling provided by Tor.

\subsection{SuperOnionBots}\label{ss:superonion}

\begin{figure}
\centering
\includegraphics[width=0.30\textwidth]{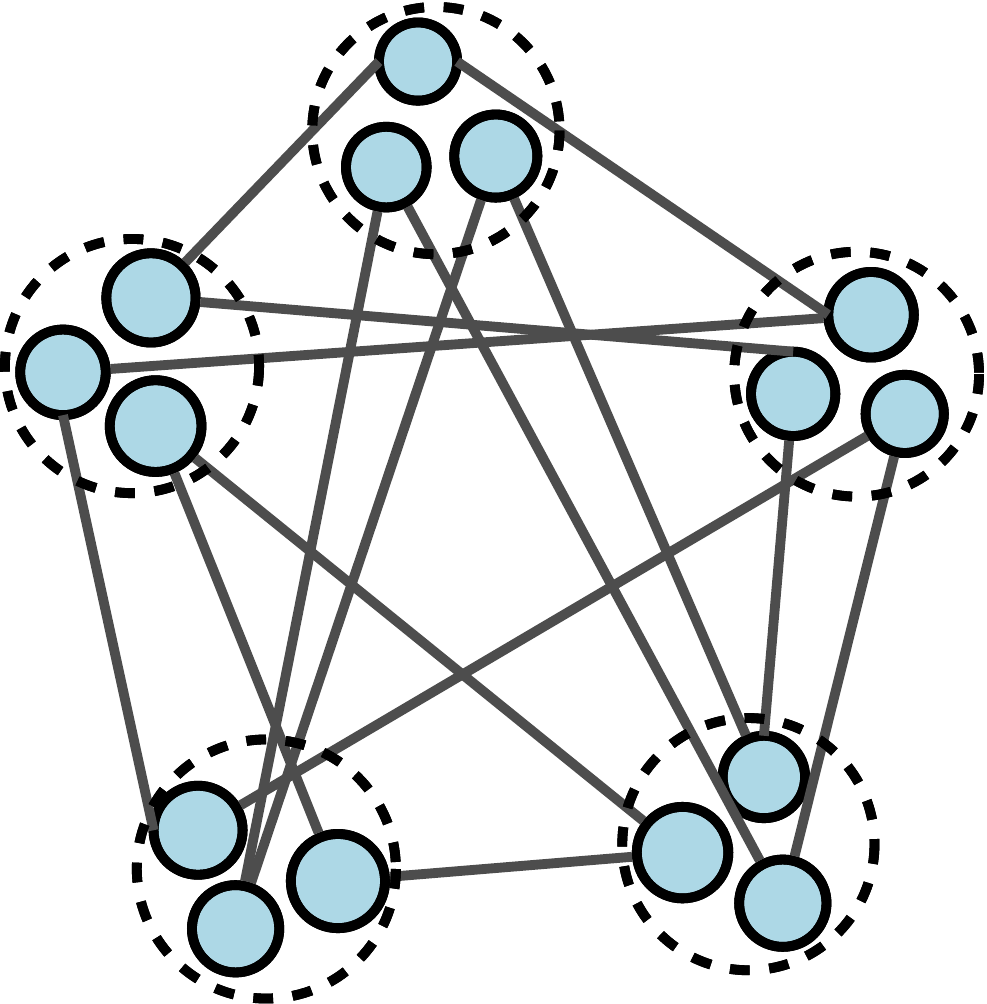}
\caption{SuperOnion construction, with $n =$ 5, $m =$ 3, $i =$ 2}
\label{f:super_node}
\end{figure}

Even though basic OnionBots exploit the decoupling provided by Tor, they do not fully utilize the decoupling of physical host, IP address and the \texttt{.onion} address. Note that a single physical host with one IP address can host many \texttt{.onion} addresses. By considering the aforementioned feature, we envision a new botnet formation called \emph{SuperOnion} (SO). This construction is consist of $n$ SOs (physical hosts) and each SO simulates $m$ virtual nodes, and each virtual node is connected to $i$ other virtual nodes; a total of $n * m$ virtual nodes, and $m *i$ virtual peers per physical node. Although in this construct each virtual nodes is still susceptible to soaping attack, the physical hosting node is immune to such attacks, as long as one of its $m$ virtual nodes is not soaped. Figure~\ref{f:super_node} illustrates a SuperOnion construction, with $n =$ 5, $m =$ 3, $i =$ 2.

To ensure the connectivity of the network, the hosting node periodically initiates a connectivity test by sending a probe message from each one of its $m$ virtual nodes, and expecting to receive the message at its $m-1$ other virtual nodes.  To maintain its stealth, the hosting node uses flooding techniques with low message complexity, such as gossiping. We make the assumption that the authorities are legally liable~\cite{Wang:2010}, and they can not participate in the botnet activity. Since the messages are encrypted and indistinguishable, the authorities are not able to drop certain message and only allow the connectivity probe messages to pass through. If certain virtual nodes don’t receive the messages, the host can deduce that they are soaped. After a physical nodes realized one of its virtual nodes is soaped, it creates a new virtual nodes, and initiates the bootstrapping stage, using peers from its currently connected virtual nodes. As we can see, distinguishing between the honest and compromised bots, translates into a variation of Byzantine generals problem~\cite{Lamport:byzantine}. We plan to investigate the operation of bots in such Byzantine environment in more detail~\cite{5654509},~\cite{Awerbuch:2002}.

\section{Related Work}\label{s:related}

In this section we look at other work that examine alternative botnet constructions. However, they still rely on traditional models, which makes them vulnerable to the current detection and mitigation, once their design is known.

Stranerger et al.~\cite{Starnberger:2008}, introduce a botnet communication protocol, called Overbot. Their design leverages Kademila peer-to-peer protocol, a distributed hash table (DHT) used by many peer-to-peer applications. They investigate the possibilities of using the existing protocol to design stealth C\&C channels. The bot uses the 160-bit hash values in a search request to announce it's sequence number, which is encrypted with the public key of the botmaster. Later this sequence number is used to send commands to the bot.

Nappa et al.~\cite{Nappa:2010}, propose a parasitic botnet protocol that exploits Skype overlay network. Skype provides a widespread resilient network with a large install base for C\&C infrastructure. The communications between the master and the bots are encrypted using adhoc schemes. The protocol broadcasts messages to all peers in the network, similar to the algorithms used in Gnutella. Once each peer receives a new message it passes it to all of its neighbors.

Vogt et al.~\cite{Vogt07armyof}, examine the possibility of creating a super-botnet by splitting a large botnet into many smaller botnets. Each smaller botnet in the super-botnet network, stores some routing information on how to communicate with the rest of the network. They use a tree-structured infection process, where each new zombie learns how many additional host it should infect and add to its botnet. This design results in a connected graph, with many densely connected cliques.

Lee and Kim~\cite{Lee:2013} explore the design and mitigation of botnets that use URL shortening services (USSes) for alias fluxing. A botmaster uses the USSes to hide and obfuscate IP address of the C\&C by using a dictionary of 256 words for each part of an IPv4. For example 10.15.43.89 can be mapped to ``Brown.Fox.Jumps.Over.'' Then this expression is transformed into a search query, such as google.com/q?=Brown+Fox+Jumps+Over. Using the URL shortening service, bots can find the corresponding IP address by using the same dictionary.

Wang et al.~\cite{WangDSN} design a hybrid peer-to-peer botnet, which is composed of servant and client bots. Their botnet communicates with a fixed number of peers contained in each bot to limit the node exposure. The botmaster can control, monitor and update the bots by sending the messages through servant bots, and getting the reports from a sensor host. These messages are encrypted using individualized predefined or dynamically generated keys.

\section{Conclusion}\label{s:conclusion}

Privacy infrastructures such as Tor had a tremendous impact on society, protecting users anonymity and rights to access information in the face of censorship. It also opened the door to abuse and illegal activities~\cite{moser2013inquiry,hopper2014}, such as ransomware~\cite{cryptolocker}, and a marketplace for drugs and contraband~\cite{biryukov2013content},~\cite{Christin:2013:silkroad}. In this work we envisioned OnionBots, and investigated the potential of subverting privacy infrastructures (e.g., Tor hidden services) for cyber attacks. We presented the design of a robust and stealthy botnet that lives symbiotically within these infrastructures to evade detection, measurement, scale estimation and observation. It is impossible for Internet Service Providers (ISP) to effectively detect and mitigate such botnet, without blocking all Tor access. Additionally, OnionBots rely on a self-healing network formation that is simple to implement, yet it has desirable features such as low diameter and low degree. Such botnets are robust to partitioning even if a large fraction of the bots is simultaneously taken down. In the scenario of a gradual take down of nodes, the network is also able to self-repair, even up to 90\% nodes deletions. More importantly, we developed soaping, a novel mitigation attack that neutralizes the basic OnionBots. We also suggested mitigations that act at the Tor level. Finally, we discussed how an attacker can design advanced botnets (SuperOnionBots) that fully utilize the decoupling provided by Tor and are immune to soaping attacks. There are still many challenges that need to be preemptively addressed by the security community, we hope that this work ignites new ideas to proactively design mitigations against the new generations of crypto-based botnets.

\bibliographystyle{IEEEtran}

\bibliography{bibliography}

\end{document}